\documentclass[reprint, amsmath,amssymb, aps, nofootinbib, superscriptaddress, floatfix, citeautoscript]{revtex4-1}
\usepackage[utf8]{inputenc}
\usepackage{graphicx}
\usepackage{dcolumn}
\usepackage{bm}
\usepackage{braket}
\usepackage{color}
\usepackage{units}
\usepackage{dsfont}
\usepackage[colorlinks, linkcolor=blue, citecolor=blue, urlcolor=blue, breaklinks=true]{hyperref}
\usepackage{float}
\usepackage{hyperref}
\usepackage{xcolor}
\usepackage{comment}

\begin{abstract}
 Wave propagation on the surface of cylinders exhibits interferometric self imaging, much like the Talbot effect in the near-field diffraction at periodic gratings. We report the experimental observation of the cylindrical Talbot carpet in weakly-guiding ring-core fibers for classical light fields. We further show that the ring-core fiber acts as a high-order optical beamsplitter for single photons, whose output can be controlled by the relative phase between the input light fields. By also demonstrating high-quality two-photon interference between indistinguishable photons sent into the ring-core fiber, our findings open the door to applications in optical telecommunications as a compact beam multiplexer as well as in quantum information processing tasks as a scalable realization of a linear optical network.
\end{abstract}

\begin{document}

\title{Talbot self-imaging and two-photon interference in ring-core fibers}

\author{Matias Eriksson}
\affiliation{Tampere University, Photonics Laboratory, Physics Unit, 33720 Tampere, Finland}

\author{Benjamin A. Stickler}
\affiliation{Faculty of Physics, University of Duisburg-Essen, 47048 Duisburg, Germany}
\affiliation{QOLS, Blackett Laboratory, Imperial College London, London SW7 2AZ, United Kingdom}

\author{Lea Kopf}
\affiliation{Tampere University, Photonics Laboratory, Physics Unit, 33720 Tampere, Finland}

\author{Markus Hiekkamäki}
\affiliation{Tampere University, Photonics Laboratory, Physics Unit, 33720 Tampere, Finland}

\author{Regina Gumenyuk}
\affiliation{Tampere University, Photonics Laboratory, Physics Unit, 33720 Tampere, Finland}

\author{Yuri Chamorovskiy}
\affiliation{Kotel’nikov Institute of Radio Engineering and Electronics (Fryazino Branch) Russian Academy of Science, 141190 Fryazino, Russia }

\author{Sven Ramelow}
\affiliation{Institut für Physik, Humboldt-Universität zu Berlin, 12489 Berlin, Germany }
\affiliation{IRIS Adlershof, Humboldt-Universität zu Berlin, 12489 Berlin, Germany}

\author{Robert Fickler}
\affiliation{Tampere University, Photonics Laboratory, Physics Unit, 33720 Tampere, Finland}

\maketitle

Near-field self-imaging of waves behind periodic grating masks, dubbed the Talbot effect \cite{talbot1836facts}, is a fundamental interference phenomenon with interesting relations to physics and math \cite{berry2001quantum}. It enables lens-free focusing, demonstrated for photons \cite{liu1989talbot,patorski1989self,case2009realization,wen2013talbot}, electrons \cite{mcmorran2009electron}, atoms \cite{chapman1995near,nowak1997high}, and even large molecules \cite{brezger2002matter, fein2019quantum}. Remarkably, this self-imaging maps onto the propagation of waves on cylindrical surfaces, where the periodicity of the angle coordinate replaces the periodic grating and where fractional Talbot revivals appear as spatially separated superpositions of the impinging wave packet \cite{niemeier1985self,baranova1998talbot,hautakorpi2006modal}.

Wave self-imaging on cylinders is caused by the discreteness of their angular momentum due to the angular periodicity of the surface \cite{robinett2004quantum}. This discreteness causes an arbitrary initial wave packet to recur at the impact angle at integer multiples of a characteristic revival length. The revival length is determined solely by the cylinder radius and the longitudinal wave number. While the wave packet remains almost uniformly dispersed between these revivals, it recurs in superpositions of localised states at fractions of the revival length. The periodicity of the surface thus acts as a linear optical element, which enables multi-port beam splitting, e.g. useful in quantum photonics applications \cite{flamini2018photonic}. Additionally, it can be applied to interferometry tasks as proposed for atomic matter-waves in torus traps \cite{kialka2020orbital}. Self-imaging on cylinders is further closely related to the orientational quantum revivals of molecular rotation states \cite{seideman1999revival,spanner2004coherent,poulsen2004nonadiabatic}, with prospects for macroscopic tests of the quantum superposition principle \cite{stickler2018probing}.

In this letter, we report the self-imaging of single photons in thin ring-core fibers. These fibers support nearly transverse orbital angular momentum modes \cite{hautakorpi2006modal,brunet2014vector}, which acquire an angular momentum-dependent phase while travelling along the fiber. At the revival length, all these phases coalesce, giving rise to the characteristic Talbot carpet, however rolled on the cylinder surface. By reconstructing the carpet, we verify that the used fibers act as high-fidelity linear optical elements generating superpositions of orbital angular momentum modes. We demonstrate that such fibers can be used to create two-fold and three-fold superpositions of the incoming light field at fractions of the characteristic revival length. Importantly, the interference pattern at the end of the fiber can be coherently controlled with additional input light fields. Moreover, we observe that the fiber can create quantum coherent superpositions of single photon states and that illuminating it with two indistinguishable photons yields the characteristic Hong-Ou-Mandel (HOM) interference \cite{hong1987measurement}. Our results demonstrate that ring-core fibers can act as compact multiplexers and de-multiplexers for optical telecommunications. Moreover, this work opens the door for using Talbot ring-core fibers as scalable higher-order beamsplitters in compact quantum optical networks.

\begin{figure*}
    \centering
    \includegraphics[width=\textwidth]{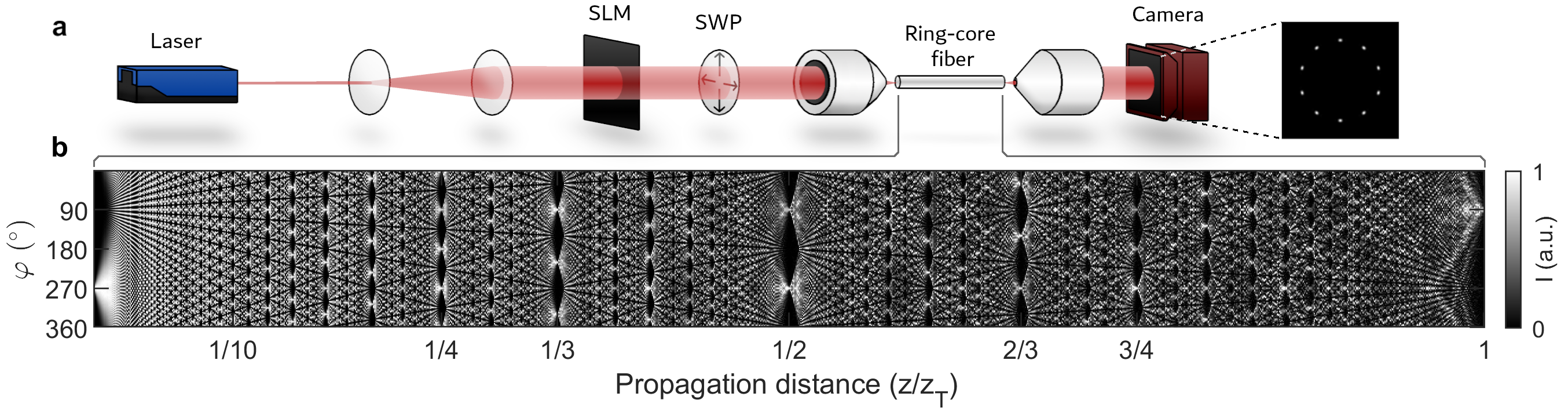}
    \caption{Sketch of the experimental setup to study the cylindrical Talbot effect. \textbf{a}, A frequency-tunable laser beam is enlarged by two lenses, modulated by a spatial light modulator (SLM), radially polarized by a S-waveplate (SWP) \cite{lai2008generation}, and focused into the ring-core fiber using a microscope objective. The output facet of the fiber is imaged using another microscope objective and a CMOS camera. The inset shows a simulated image at one-tenth Talbot length $z_{\rm T}/10$. \textbf{b}, Simulated intensity distribution of light propagating in a ring-core fiber, i.e. the cylindrical Talbot carpet, with a single Gaussian input spot. The intensity is recorded at the central radius of the ring-core, and normalized for each simulation step.}
    \label{fig1}
\end{figure*}

The experimental setup to study the self-imaging effect is sketched in Fig.\,\ref{fig1} (a).
The ring-core fiber is placed on a flat metal plate. A laser with frequency $\omega$ is focused onto an approximately Gaussian spot at the input facet of the fiber ring. The light propagates along the fiber and is detected with a camera after leaving the fiber. We use custom-fabricated ring-core fibers with an inner diameter of $2R = 54.7~\mu$m and a ring width of $w = 2.15~\mu$m. The fabrication of the fibers is done by a standard, multi-step process described in more detail in the supplementary material \cite{supp}. The ratio of the ring diameter along the two transverse axes differs from unity for a perfect circular shape by less than one percent.
The cladding material of the fiber is pure silica glass with a refractive index of $n_{\rm o}$=1.454 for a wavelength of 780~nm \cite{MalitsonI.H1965ICot}, and the refractive index in the ring-core is $n_{\rm i}$=1.467 (see supplementary material for more information \cite{supp}).

The small difference between the refractive indices implies $\Delta n/n_{\rm o} \simeq 9\times 10^{-3}\ll 1$, so that the light propagation in the ring is well described by Maxwell's equations in the weak-guiding approximation \cite{gloge1971weakly}. In this regime, the propagating modes in the fiber are linearly polarized ${\rm LP}_{p\ell}$ modes characterised by a radial mode index $p\geq 0$ and the orbital angular momentum index $\ell \in \mathbb{Z}$ \cite{sarkar2001analysis,marcou2003comments}. For small ring widths, $w/R\ll 1$, only the lowest radial modes contribute to the dynamics and the longitudinal propagation constant $\beta \simeq k - \ell^2/2 k R^2$ depends quadratically on the angular momentum $\ell$, where $k = \omega n_{\rm i}/c$ is the wave-vector in the ring. The resulting wave amplitude at distance $z$ and angle $\varphi$ on the cylinder surface follows from the wave equation in the paraxial approximation as \cite{supp}
\begin{equation}\label{eq:scalwave}
    \psi(\varphi,z) = \sum_{\ell \in \mathbb{Z}} \psi_\ell  \, e^{i \ell \varphi} \exp \left (ik z -i \frac{z \ell^2}{2k R^2} \right ),
\end{equation}
where $\psi_\ell$ are the Fourier components of the field at the fiber input port $z = 0$.

\begin{figure}
    \centering
    \includegraphics[width=0.95\linewidth]{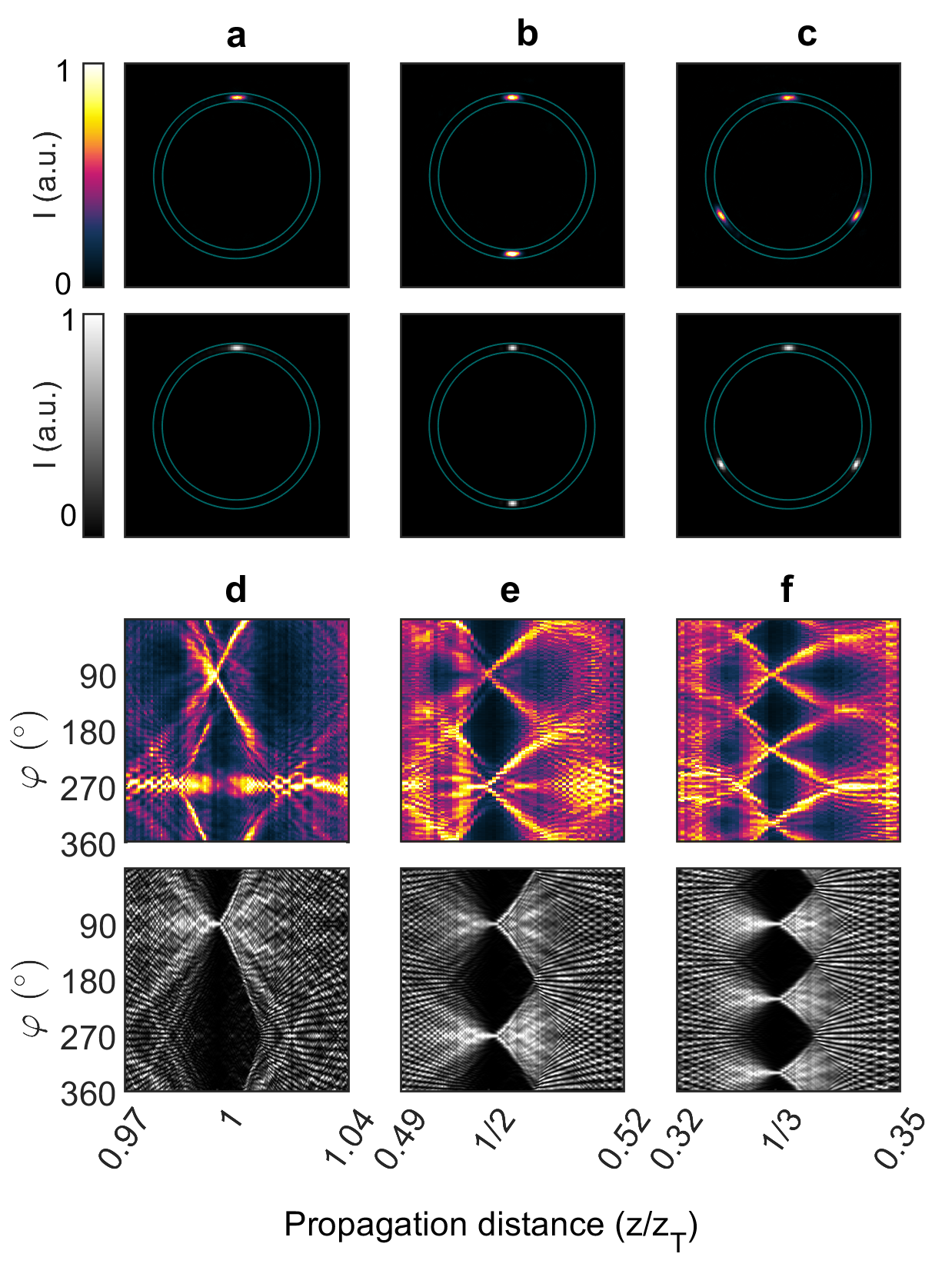}
    \caption{{Experimental and simulated intensity distributions of output light fields for different fractional Talbot lengths of the ring-core fiber and a single Gaussian spot as the input field.} 
    \textbf{a-c}, Intensity distributions at the output facet (depicted by thin teal lines) for fiber lengths of $z_{\rm T}$ (a), $z_{\rm T}/2$ (b), and $z_{\rm T}/3$ (c), recorded with a camera (upper row) and simulated (lower row). 
    \textbf{d-f}, Parts of the cylindrical Talbot carpet around $z_{\rm T}$ (d), $z_{\rm T}/2$ (e), and $z_{\rm T}/3$ (f), generated from camera images while scanning the wavelength (upper row) and simulations (lower row).}
    \label{fig2}
\end{figure}

The light field \eqref{eq:scalwave} exhibits self-imaging at the opposite side of the cylinder if the longitudinal position is an integer multiple of the Talbot length
\begin{equation} \label{eq:talbot}
z_{\rm T} = 2 \pi k R^2,
\end{equation}
and all phases coalesce so that $\psi(z_{\rm T},\varphi) = \psi(0,\varphi+\pi)$, as shown in Fig.\,\ref{fig1} (b). In addition, at fractions of the Talbot length, the initial wave packet reappears in a balanced superposition at different angles. For instance, at one-half and at one-third Talbot length the wave re-appears in a superposition of two and three localised wave packets, respectively. 

\begin{figure*}
    \centering
    \includegraphics[width=\linewidth]{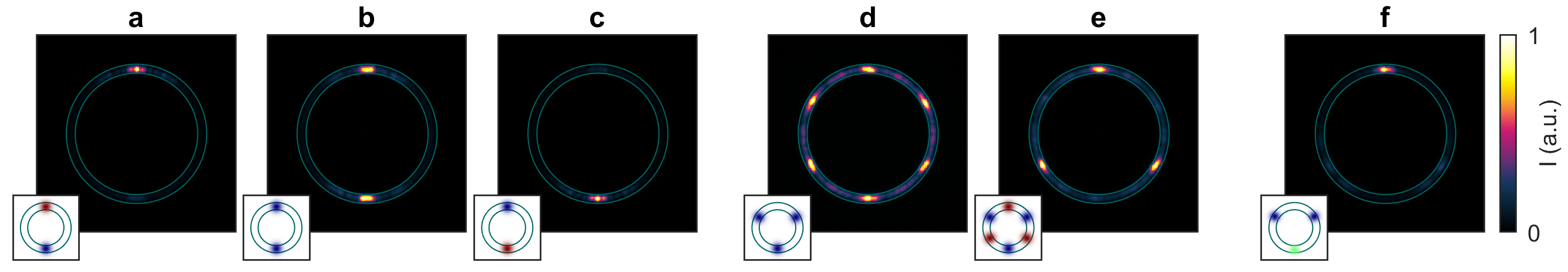}
    \caption{{Experimental intensity distributions of the output fields of ring-core fibers with multiple Gaussian input spots with varying relative phases.} The input light fields are described by the insets, where the colors denote relative phases of 0 (blue), $\pi/2$ (red), and $2\pi/3$ (green). \textbf{a-e}, Output fields at half Talbot length, demonstrating a tunable 2-input 2-output coupler (a-c), and three simultaneous beamsplitters (d) and beam combiners (e). \textbf{f}, Output field at one-third Talbot length showing a tritter used as a beam combiner with three input spots.}
    \label{fig3}
\end{figure*}

Using standard cleaving techniques, we cut the ring-core fiber into multiple pieces with a precision of around 0.5~mm. The fiber pieces match a full, a half, and a third Talbot length, which corresponds to lengths of 6.00~cm, 3.00~cm and 2.00~cm, respectively, for the chosen vacuum wavelength of 780~nm.
Light fields launched into these fibers at single input positions revive at a single, two, and three output positions, respectively, as shown in Fig.\,\ref{fig2} (a-c) in the upper row.
The experimental data is also compared with numerical simulations of the scalar field dynamics in the weak-guiding approximation using the split-step propagation method \cite{poon2017engineering}. In the simulations, light with a transverse Gaussian beam profile and a beam waist of $w_0$=1.5~$\mu$m is launched into a ring-core with dimensions matching the experimental specifications. The propagation is simulated with propagation steps of 0.1~$\mu$m and absorbing boundaries to account for losses.
The experimental data shows excellent agreement with the simulations, see Fig.\,\ref{fig2} (a-c), demonstrating that ring-core fibers can act as high-order beamsplitters with high fidelity.

Resolving the revivals of the input field requires cleaving the fiber with a precision exceeding what standard cleaving techniques allow. In order to circumvent this limitation, we utilize the wavelength dependence of the Talbot length, see Eq.~\eqref{eq:talbot}, and tune the wavelength of the laser to match the desired fractions of the Talbot length exactly to the lengths of the fibers. 
Moreover, by using this wavelength dependence, it is possible to map the carpet in the vicinity of the full, half, and third revivals. As can be seen in Fig.\,\ref{fig2} (d-f), the obtained patterns closely follow the distinct patterns obtained in simulations of light propagating through the fiber.
The self-imaging effect is perfectly visible with only slight degradation for the different fractional Talbot lengths. The degradations result from accumulated imperfections after long distances and a small ellipticity of the ring-core fiber (see \cite{supp} for more information). 
We note that the strong length-dependence of the Talbot effect might be exploited for high-precision length measurements of ring-core fibers well beyond that of a physical ruler.

To demonstrate the coherence of the generated superpositions at the half and third Talbot length, we shape the input field to consist of multiple input spots with well-controlled relative phases using a spatial light modulator (SLM) \cite{bolduc2013exact}, and observe their interference at the fiber output, see Fig.\,\ref{fig3}. To additionally verify that the output pattern appears independent of the angular input position, the experimental results are averaged from angle-corrected output light fields of 60 evenly distributed angular input positions. This method is additionally used to show that the patterns shown in Fig.\,\ref{fig2} (a-c) also appear independent of the angular input position, with the results shown in the supplementary material \cite{supp}.

In the case of two inputs, we illuminate the ring-core with two Gaussian spots on opposite sides of the ring with a variable relative phase thereby demonstrating that the self-imaging phenomenon can be used as a 2-input 2-output coupler similar to a standard beamsplitter. For a relative phase of $\pi/2$, the two input light fields constructively interfere in the fiber, such that at the output facet light only appears at a single angular region equal to one of the input positions [see Fig.\,\ref{fig3}~(a)]. 
Furthermore, by varying the relative phase of the two input spots, the interference can be continuously tuned from constructive to destructive at both of the two possible output spots. This is demonstrated in Fig.\,\ref{fig3} (b) and (c), presenting two distinct examples of no interference (two output spots) and constructive interference on opposite sides, respectively.

The tunability of the Talbot interference with a relative phase between several inputs can be used to determine and compensate for an overall tilt of the fiber with respect to the beam axis. Compensation of this tilt enables us to simultaneously implement multiple beamsplitters using a single ring-core fiber. Here, each beamsplitter corresponds to the input-output regions along a certain transverse axis and can be independently addressed. In Fig.\,\ref{fig3} (d) and (e), we show two examples of three simultaneous beamsplitters that are configured to perform a beam splitting [Fig.\,\ref{fig3} (d)] and combining [Fig.\,\ref{fig3} (e)], respectively.

The fractional Talbot revivals can also be used as higher-order beamsplitters when the fiber lengths match smaller fractions of the Talbot length. For a fiber of one-third Talbot length, the Talbot effect corresponds to a so-called tritter operation \cite{weihs1996all}. When sending three coherent input light fields at equally spaced positions with a specific phase relation into the ring-core, they interfere into a single output beam as shown in Fig.\,\ref{fig3} (f). 

\begin{figure}
    \centering
    \includegraphics[width=\linewidth]{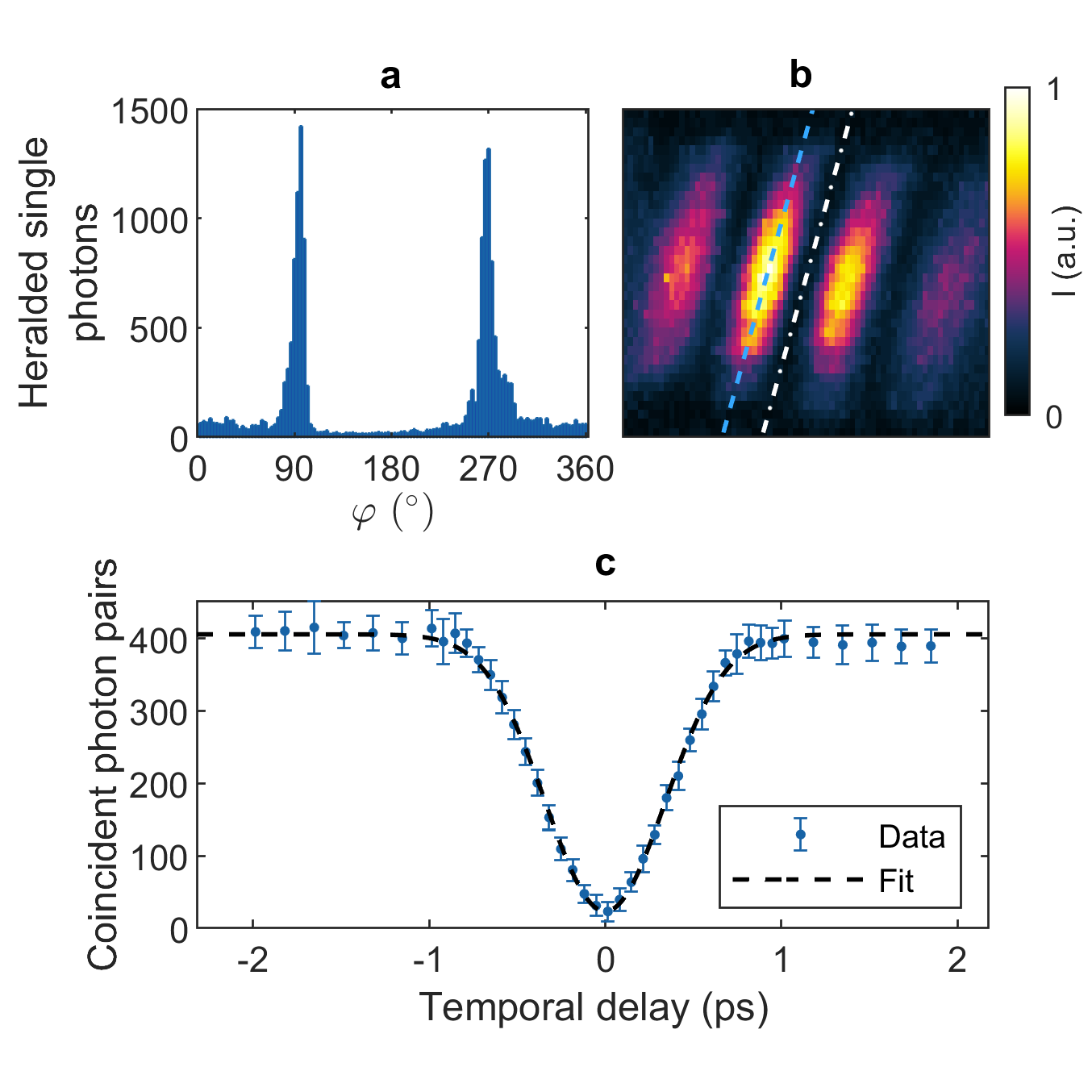}
    \caption{{Cylindrical Talbot effect using single photons.} \textbf{a}, Angular distribution of heralded single photons of the output of a half Talbot length fiber with a single Gaussian input spot, measured over 20~s for each angular position. \textbf{b}, Interference fringe pattern generated by overlapping the two spots of the output light field described in (a), captured using unheralded photons from the single photon source and a high-sensitivity camera. The two parallel dotted lines show the positions of the slit, which were used to determine the visibility of the fringes to be around 96~\% using heralded single photons. \textbf{c}, HOM interference curve, generated by scanning the temporal delay between two indistinguishable photons focused onto opposite sides of the input facet of a half Talbot length fiber. The coincident photon pairs are measured at each delay position for 3 seconds. The error bars are standard errors calculated from 20 consecutive repetitions of each measurement, and the dashed curve is a fit.}
    \label{fig4}
\end{figure}

Finally, we show that the cylindrical Talbot effect implements a quantum coherent beamsplitter for single photons.
In order to demonstrate this, we measure single-photon interference using heralded single photons and two-photon interference using degenerate  photon pairs (see \cite{supp,bouchard2018experimental} for details), 
thus also showing that a broader bandwidth of up to 3~nm does not hinder the implementation of the complex self-imaging and interferences along the ring-core fiber. 
This paves the way for a future use of the effect in quantum optics experiments.
To this end, we focus the heralded single photons onto a single spot on the ring-core, and propagate them through a fiber of half Talbot length.
To spatially resolve the resulting output pattern, we implement an opaque mask, which is only transmittive for an angular region of around 3~degrees. By scanning the mask over a full rotation and detecting the transmitted photons using a single photon bucket detector, we obtain an output pattern that nicely shows two output spots on opposing sides, see Fig.\,\ref{fig4} (a). To show that the two outputs are quantum coherent, we interfere the output photon using a lens with a focal distance of 300~mm leading to a distinct fringe pattern shown in Fig.\,\ref{fig4} (b) in the focal plane. We place a narrow slit ($\approx$ 25~$\mu$m) in the global maximum and the neighbouring minimum of the generated interference pattern, as indicated in Fig.\,\ref{fig4} (b), and record the transmitted photons using a bucket detector. We observe high contrast interference fringes with a visibility of 96$(^{+3}_{-8})$~\%. 

In addition to verifying single-photon interference, we measure two-photon interference using the fiber.
To achieve two-photon bunching, namely HOM interference \cite{hong1987measurement}, we focus two indistinguishable photons onto opposite sides of the input facet of a ring-core fiber, cut to half Talbot length.
We then vary their temporal distinguishability by delaying one of the photons with an adjustable delay line. As can be seen in Fig.\,\ref{fig4} (c), when the delay is set to zero, the photons bunch and exit the fiber from the same position on the ring. Consequently, the number of coincident photon detections from opposite sides of the ring-core decrease significantly.
From a fit to the measured data we calculate a visibility of $V = (R_{cl}-R_{qu})/R_{cl} = {94\pm 1 }~\%$ for the HOM curve, which is well-beyond the classical limit of 50~\%, and where the error corresponds to the standard error.
In the visibility, $R_{cl}$ denotes the classically expected rate without any bunching and $R_{qu}$ is the rate observed due to photon bunching.
Hence, the observed self-imaging effect in ring-core fibers can split photons into spatially separated, coherent outputs  while enabling multi-photon interference effects.
As such, the Talbot effect might be useful in the future for quantum photonic processing tasks, e.g. as a compact scalable linear optical network \cite{flamini2018photonic}

In conclusion, the fractional Talbot effect in thin ring-core fibers provides a promising platform for simple and scalable realizations of quantum coherent higher-order input-output couplers. 
The non-trivial topology of the waveguide replaces the grating in the standard Talbot self-imaging, enabling beam-splitting and interference with no diffractive elements.
It can be used as a convenient way to realize a multi-input multi-output device, which scales favorably as it becomes more compact for higher-order realizations. 
From simulations we estimate that by using our system at least 40 distinct input and output ports can be realized with a fiber length of around 1.5~mm.
In the quantum domain, the cylindrical Talbot effect in ring-core fibers can serve as a novel tool to perform quantum computation and simulation tasks, for example as an high-efficiency realization of multi-mode unitary manipulations \cite{carolan2015universal,flamini2018photonic,leedumrongwatthanakun2020programmable, brandt2020high, hiekkamaki2021high}.
For both classical and quantum domains, studying ways to tune the effect, e.g. through local heaters acting as thermo-optic phase modulators \cite{carolan2015universal}, will be an important effort for future developments.
Finally, as the effect is sensitive to length and wavelength in a well defined manner, it might also find applications as a novel type of sensor of either. 

{\it Acknowledgment ---}  The authors thank S. Prabhakar, T. Niemi, M. Närhi and S. Bej for fruitful discussions. The authors further thank I. Vorob'ev, V. Aksenov, A. Kolosovskiy and V. Voloshin for the fiber manufacturing.
ME, LK, MH, and RF acknowledge the support of the Academy of Finland (Grant No. 308596) and the support of the Academy of Finland through the Competitive Funding to Strengthen University Research Profiles (Decision 301820).
ME, LK, MH, RG, and RF  acknowledge the support from the Flagship of Photonics Research and Innovation (PREIN) funded by the Academy of Finland (Grant No. 320165). 
LK acknowledges support from the Vilho, Yrjö and Kalle Väisälä Foundation of the Finnish Academy of Science and Letters. 
MH acknowledges support from the Doctoral School of Tampere University and the Magnus Ehrnrooth foundation through its graduate student scholarship.
SR acknowledges support from Deutsche Forschungsgemeinschaft (RA 2842/1-1).
RF acknowledges support from the Academy of Finland through the Academy Research Fellowship (Decision 332399). 

\bibliography{references}

\end{document}


\title{Supplemental Material:  Talbot self-imaging and two-photon interference in ring-core fibers}

\author{Matias Eriksson}
\affiliation{Tampere University, Photonics Laboratory, Physics Unit, 33720 Tampere, Finland}

\author{Benjamin A. Stickler}
\affiliation{Faculty of Physics, University of Duisburg-Essen, 47048 Duisburg, Germany}
\affiliation{QOLS, Blackett Laboratory, Imperial College London, London SW7 2AZ, United Kingdom}

\author{Lea Kopf}
\affiliation{Tampere University, Photonics Laboratory, Physics Unit, 33720 Tampere, Finland}

\author{Markus Hiekkamäki}
\affiliation{Tampere University, Photonics Laboratory, Physics Unit, 33720 Tampere, Finland}

\author{Regina Gumenyuk}
\affiliation{Tampere University, Photonics Laboratory, Physics Unit, 33720 Tampere, Finland}

\author{Yuri Chamorovskiy}
\affiliation{Kotel’nikov Institute of Radio Engineering and Electronics (Fryazino Branch) Russian Academy of Science, 141190 Fryazino, Russia }

\author{Sven Ramelow}
\affiliation{Institut für Physik, Humboldt-Universität zu Berlin, 12489 Berlin, Germany }
\affiliation{IRIS Adlershof, Humboldt-Universität zu Berlin, 12489 Berlin, Germany}

\author{Robert Fickler}
\affiliation{Tampere University, Photonics Laboratory, Physics Unit, 33720 Tampere, Finland}

\maketitle
\section{Cylindrical Talbot effect}

The cylindrical Talbot effect follows from the scalar wave equation in a cylindrical shell of radius $R$,
\begin{equation}
    \left ( \partial_z^2 + \frac{1}{R^2} \partial_\varphi^2  - k^2 \right ) \psi(\varphi,z) = 0.
\end{equation}
Here, $k$ is the wavenumber in the cylindrical shell, $z$ is the waveguide symmetry axis and $\varphi$ denotes the azimuthal angle. In the paraxial regime of small transverse momentum, we set $\psi(\varphi,z,t) \simeq \phi(\varphi,z) e^{i k z}$. The term $\vert \partial_z^2 \phi\vert \ll 2 k |\partial_z \phi|$ can then be neglected, yielding
\begin{equation}
    2 i k \partial_z \phi = - \frac{1}{R^2}\partial_\varphi^2 \phi.
\end{equation}
This equation can be solved exactly by Fourier expanding the angular coordinate in orbital angular momentum modes,
\begin{equation}\label{eq:scalwave}
    \phi(\varphi,z) = \sum_{m \in \mathbb{Z}} \psi_m e^{i m\varphi} \exp \left ( -i \frac{z m^2}{2k R^2} \right ),
\end{equation}
where $\psi_m$ are the Fourier coefficients of the incoming transverse wave amplitude $\phi_0(\varphi) = \phi(\varphi,z = 0)$.

The scalar wave \eqref{eq:scalwave} is periodic in $z$ with period $2 z_{\rm T}$, where $z_{\rm T} = 2 \pi k R^2$ is the cylindrical Talbot length. Between $z = 0$ and $z = z_T$, Eq.\,\eqref{eq:scalwave} describes the Talbot carpet wrapped around the cylinder.

Specifically, at the Talbot length, the impinging wavepacket re-appears at the opposite side of the ring,
\begin{subequations}
\begin{equation}
    \phi(\varphi,z_{\rm T}) = \phi_0(\varphi + \pi),
\end{equation}
while the incoming wave appears in a superposition of shifted copies at fractions of the Talbot length, e.g.
\begin{equation}
    \phi(\varphi,z_{\rm T}/2) = \frac{e^{-i\pi/4}}{\sqrt{2}} \left [ \phi_0(\varphi) + i  \phi_0(\varphi + \pi) \right ],
\end{equation}
\begin{align}
    \phi(\varphi,z_{\rm T}/3) = & \frac{1}{\sqrt{3}} \left [ e^{-i \pi/6} \phi_0(\varphi + \pi/3) + i \phi_0(\varphi + \pi) \right. \nonumber \\
     & \left. +e^{-i \pi/6} \phi_0(\varphi + 5\pi/3)\right ],
\end{align}
and
\begin{align}
    \phi(\varphi,2z_{\rm T}/3) = & \frac{1}{\sqrt{3}} \left [ e^{i \pi/6} \phi_0(\varphi + 2\pi/3) - i \phi_0(\varphi) \right. \nonumber \\
     & \left. +e^{ i\pi/6} \phi_0(\varphi + 4\pi/3)\right ].
\end{align}
\end{subequations}

\section{Ring-core fiber}

\subsection{Fabrication}
The `mother' preform for the ring core fiber was made by the common modified chemical vapor deposition (MCVD) method with depositing different doped silica layers on an internal surface of a support silica tube. First layers with GeO$_\text{2}$-doping were made forming the ring core region. After this, several pure silica SiO$_\text{2}$-layers were deposited forming a protection cladding. To complete the mother preform the pure silica rod was inserted into a support tube with deposited layers and this assembled structure was collapsed in a solid rod preform at a very high temperature of around 2000$^\circ$ Celsius. Finally this preform was drawn into a fiber with the required diameter and a polymer acrylate coating was applied to protect the fiber surface.

\subsection{Characterization}
\begin{figure}
    \centering
    \includegraphics[width=\linewidth]{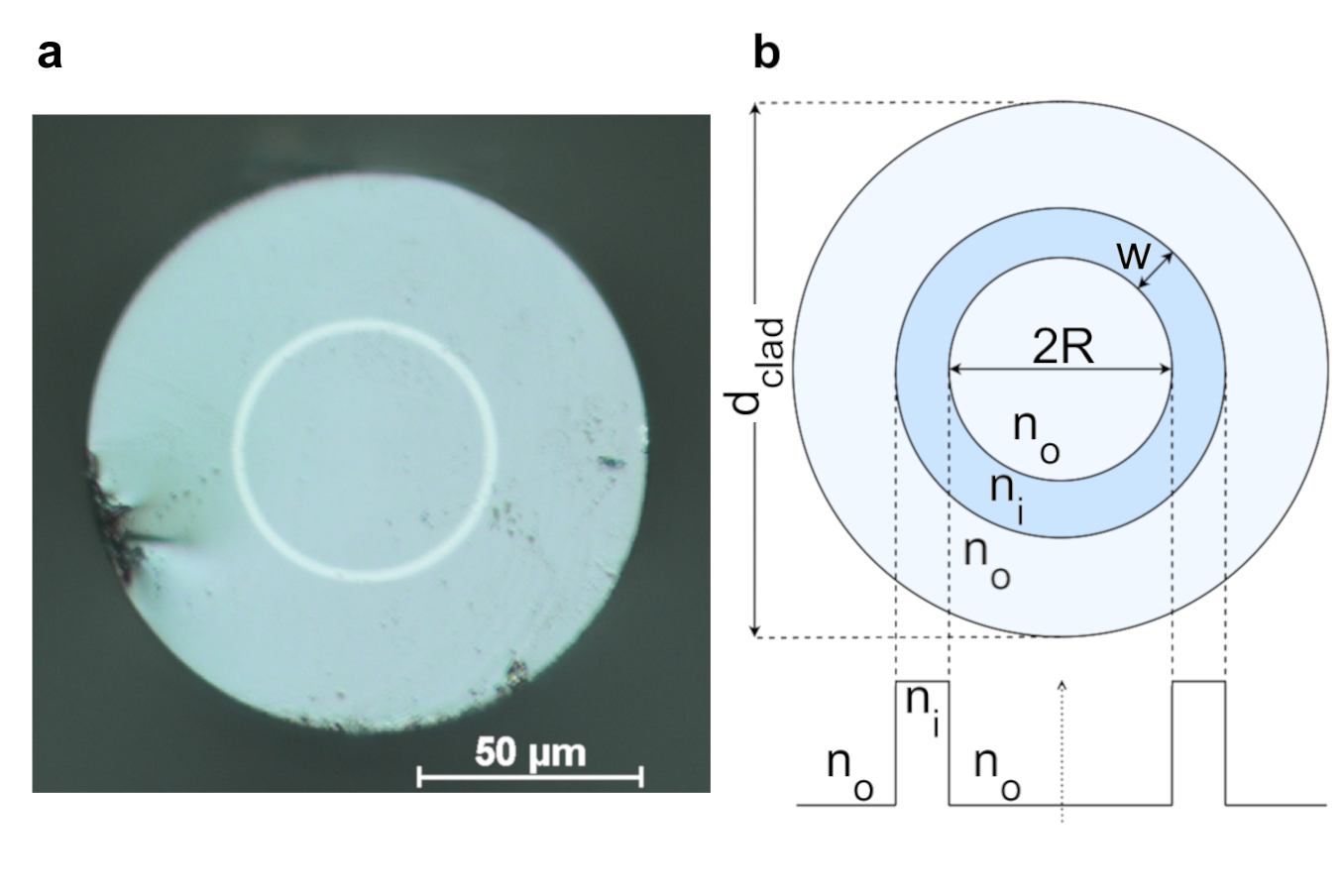}
    \caption{\textbf{Ring-core fiber:} \textbf{a}, Microscope image of the ring-core fiber. \textbf{b}, Sketch of the index profile of the utilized fiber (not to scale).}
    \label{fig:fiber}
\end{figure}

To inspect the fabricated ring-core fiber, we image the end facets of cleaved fiber pieces using 20x, 50x, and 100x microscope objectives. An image of a fiber facet is shown in Fig.\,\ref{fig:fiber} (a). By averaging values obtained from eight images, we obtain inner ring-core diameters of $a$ = 55.14~$\mu$m and $b$ = 54.61~$\mu$m for the two major axes. Hence, the ring-core fiber has an ellipticity $e$ = $\sqrt{1-(b/a)^2}$ of around $e$ $\approx$ 14~\%. 
In Fig.\,\ref{fig:fiber} (b), we show a sketch of the index profile of the fiber. The relative refractive index distribution is measured before drawing the fiber using a preform analyser (Photon kinetics, P-101 Preform analyser). The cladding material is pure silica glass with a refractive index of $n_0$ = 1.454 for a wavelength of 780~nm \cite{MalitsonI.H1965ICot}, which, together with the relative refractive index distribution gives the value $n_i$ = 1.467 as the refractive index of the ring-core.

\section{Experimental Setup}

\subsection{Setup}
The setup depicted in Fig.\,1 (b) of the main text  enables the flexible shaping of the input light field and the observation of the self-imaging effect at the output of the ring-core fiber.
At first, the beam from a tunable single-frequency laser (Toptica, DL pro, $<$ 1~MHz linewidth) is enlarged and illuminates the screen of a spatial light modulator (SLM, Holoeye Pluto, 1920x1080 pixels) in a nearly uniform way, i.e. an approximate plane wave light field.
Using amplitude and phase modulation techniques implemented by computer-generated holograms \cite{bolduc2013exact}, the required light field at the desired position is carved out of the plane wave.
Although being lossy, this technique enables easy adjustments of the angular position of the input light as well as the launching of more complex input light fields into the fiber by simply reprogramming the displayed hologram.
As the effect of self-imaging works better for a radially polarized light field (see comparison in Fig.\, \ref{fig:ComparePol}), an S-waveplate \cite{lai2008generation} is placed in the beam path before the fiber. 
The S-waveplate can be seen as a half waveplate with an azimuthally varying optical axis, such that the uniform linear polarization of the light required by the SLM is transformed to be radially polarized.
Using a lens system (not shown in Fig.\,1 of the main text) and a microscope objective, the modulated light field is demagnified and reimaged onto the input facet of the ring-core fiber such that it illuminates one (or multiple) angular position with a Gaussian spot size of around $\omega_0$=1.5~$\mu$m.
The respective fiber piece is placed on a flat metal plate. 
In order to minimize the stress induced by clamping or a possible bending of the fiber, no external holder is used to keep the fiber in place. 
After the light is transmitted through the fiber, the output facet of the fiber is magnified and imaged using a second microscope objective such that the obtained pattern can be recorded with a CMOS camera (ZWO ASI120MM Mini). 

\subsection{Heralded single photon source} 

To verify that the Talbot-effect in ring-core geometries also works for single photons, we use a heralded single photon source based on the nonlinear optical process of spontaneous parametric down conversion.
Using a nonlinear crystal (30~mm length, ppKTP crystal, type~0) and a single frequency pump laser (523.7~nm wavelength), we generate photon pairs with a heralding photon with a wavelength of around 1585~nm detected by a single-photon nanowire detector and a signal photon at 782~nm detected by a single photon avalanche photodiode (Laser Components COUNT T).
The heralding photons are then correlated with the heralding photons using a time tagger (IDS ID900).
With a pump power of 13~mW we obtain around 50~kHz heralding events after coupling both photons into single-mode fibers, and, as a quality-measure, a $g^{(2)}(0)$-value of 0.24 $\pm$ 0.03 \cite{bouchard2018experimental}.
When recording the unheralded photons that propagate through the ring-core fiber as displayed in Fig.\,4 (b) of the main text, we use a high-efficiency CMOS camera (ZWO ASI1600 Pro).

\subsection{Degenerate photon pair source}

In order to measure two-photon interference, we use a two-photon source producing degenerate photon pairs. 
We achieve this, again, through spontaneous parametric down conversion by pumping a 12~mm long, type~0 ppKTP crystal with a continuous wave laser with a wavelength of 405~nm.
The free-space power of the pump laser is 134~mW with a linewidth below 110~GHz.
The down-converted photons are filtered through a 3~nm bandpass filter, with a center wavelength of 810~nm.
The degenerate photon pair is then split by using its momentum correlation and coupled to single mode fibers (SMFs), effectively filtering each photon to a single transverse-spatial mode.
Before coupling the photons into SMFs, one of the photons goes through a delay line which is implemented by placing two mirrors onto a computer-controlled translation stage.
This stage is then moved to make the photon pair temporally indistinguishable in the ring-core fiber.
From this photon source, we get roughly 2.7~MHz of accidental-corrected photon pairs coupled into the SMFs.
Right after the source, the visibility of the two-photon interference is $V = 97.0\pm 0.1~\%$ in a fiber beamsplitter (Thorlabs TW805R5F2).  The errors are standard errors calculated from a fit to the two-photon interference data \textbf{\cite{hiekkamaki2021high}}.
After passing through the ring-core fiber, the photon pairs are detected using two single photon avalanche photodiodes (Laser Components COUNT T) and a time tagger (IDS ID900).

\section{Additional measurements and simulations}

\subsection{Single input light fields}
\begin{figure}
    \centering
    \includegraphics[width=\linewidth]{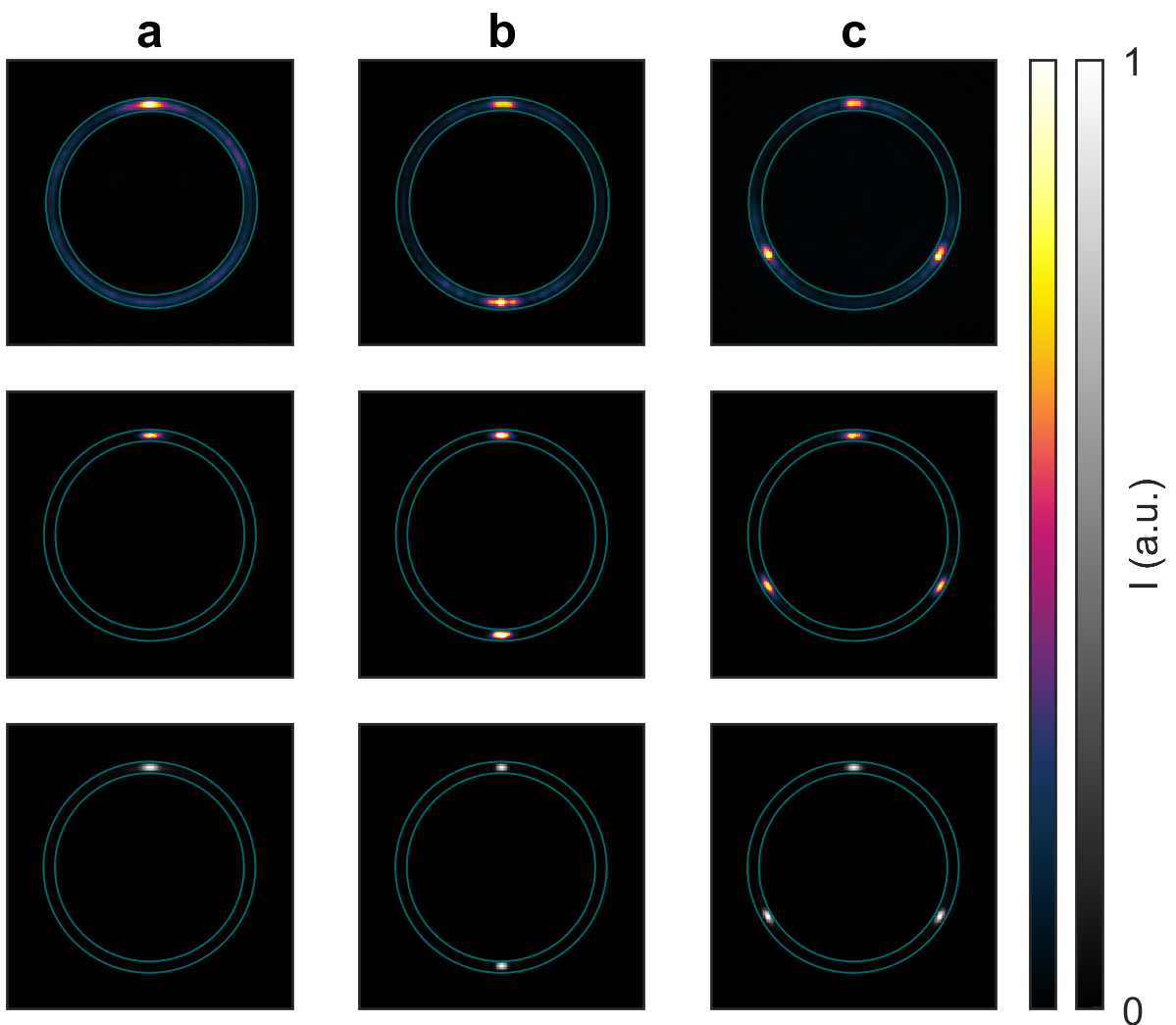}
    \caption{\textbf{Output intensity patterns using a single input light field:} The intensities shown correspond to ring-core fibers of a full (\textbf{a}), half (\textbf{b}), and third (\textbf{c}) Talbot length. The upper row shows the angle-corrected averaged intensity for 60 different angular input positions of the light field shaped by a spatial light modulator. The middle row shows the output intensity when the input is an (unmodulated) tightly focused Gaussian beam launched into the bottom of the ring core fiber. The lower row corresponds to the simulated intensities.}
    \label{fig:CompareSLMvsDirect}
\end{figure}
In the main manuscript, we show the results obtained by illuminating the ring-core fibers of three different lengths with a single input spot using a microscope objective and compare the obtained output patterns with the ones obtained in simulations.
The results are shown in the main manuscript in Fig.\,2 (a-c) and in Fig.\,\ref{fig:CompareSLMvsDirect} in the lower two rows.
In contrast, here we perform the same measurements, but using a spatial light modulator (SLM) to shape an enlarged Gaussian beam (as shown in the setup in Fig.\,1 in the main text) using holographic amplitude and phase modulation \cite{bolduc2013exact}.
By reprogramming the holographic modulation we launch the light into the ring-core fiber at 60 different angular input positions to test that a similar pattern appears independent of the exact angular position. The average over all output patterns is shown in the top row of Fig.\,\ref{fig:CompareSLMvsDirect}. The expected output is still perfectly visible, which nicely demonstrates the rotational symmetry of the system.
The quality of the output fields with an input light field modulated by an SLM are slightly degraded compared to the case where the laser is directly focused on the ring-core, omitting the SLM. We attribute this to small imperfections in the holographic modulation scheme used in combination with the SLM screen, which also introduces slight distortions to the modulated light field.

\subsection{Effect of ellipticity of the ring-core fiber}
\begin{figure}
    \centering
    \includegraphics[width=\linewidth]{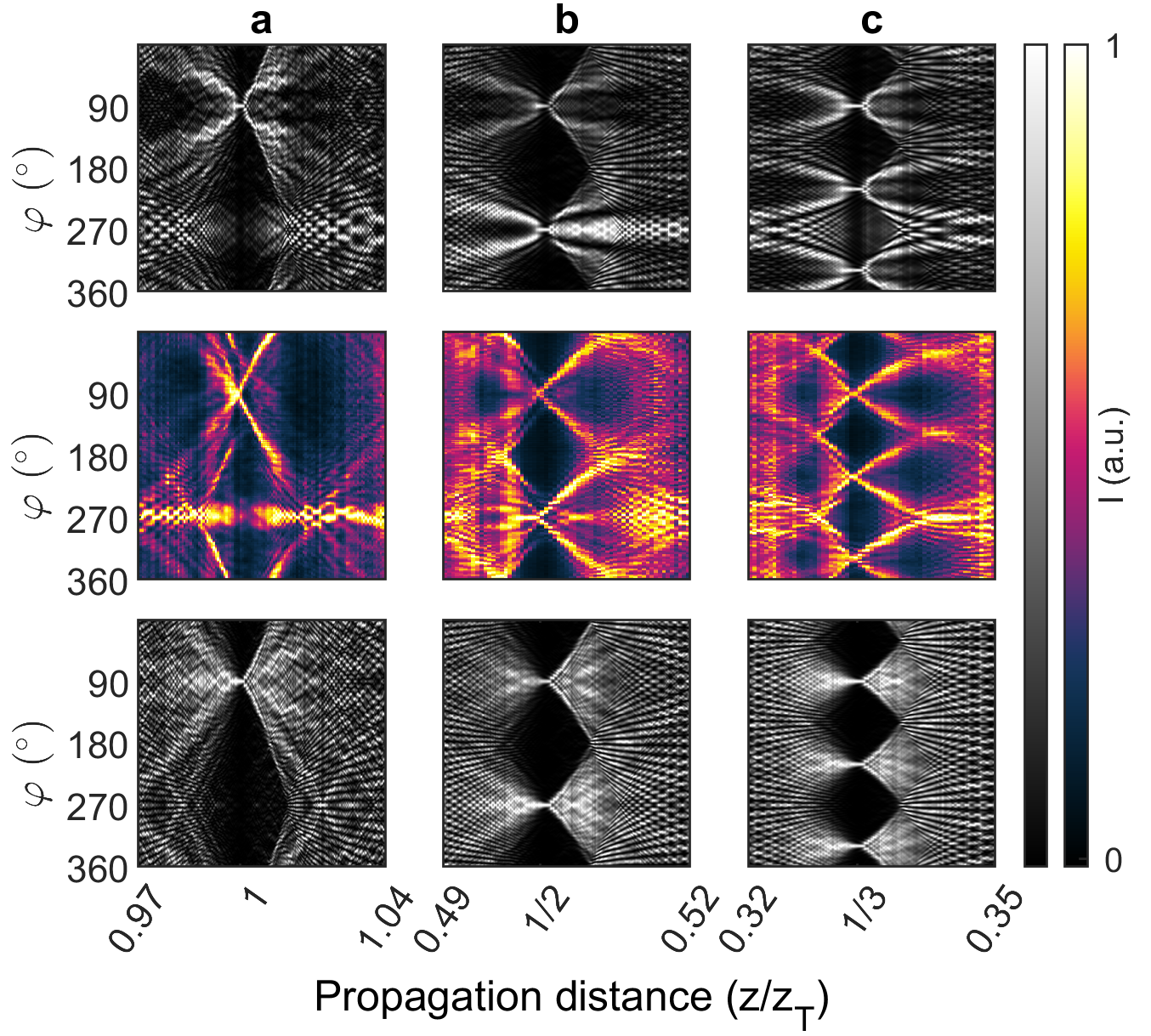}
    \caption{\textbf{Study of an elliptical ring-core fiber:} The intensity carpets shown correspond to ring-core fibers of a full (a), half (b), and third (c) Talbot length. The upper row shows a simulation for a fiber with an ellipticity similar to the ellipticity measured for the fibers used in the experiments. The middle row shows the experimentally obtained carpet. In the lower row, the simulations for a perfectly circular ring-core are shown.}
    \label{fig:CompareEllip}
\end{figure}
As determined from microscope images of the cleaved fiber pieces, the ring-core of the utilized fibers have a slight ellipticity of around 14~\%. 
To observe the effects of the ellipticity of the ring-core, we simulate the propagation through such a deformed ring-core fiber and record the Talbot carpet.
The simulated results for the elliptical fiber shown in the upper row of Fig.\,\ref{fig:CompareEllip} exhibit slight deformations compared to the results obtained for a perfectly circular ring-core.
However, the experimental results obtained by scanning the wavelength and recording the output intensities also exhibit similar deformations, matching better to the simulations using the elliptical fiber rather than a perfectly circular one.
For comparison, we display the experimental data and simulations with a perfectly circular ring-core in the main text in Fig.\,2 (d-f), and also in Fig.\,\ref{fig:CompareEllip} in the lower two rows.

\subsection{Different polarizations of the input light field}
\begin{figure}
    \centering
    \includegraphics[width=\linewidth]{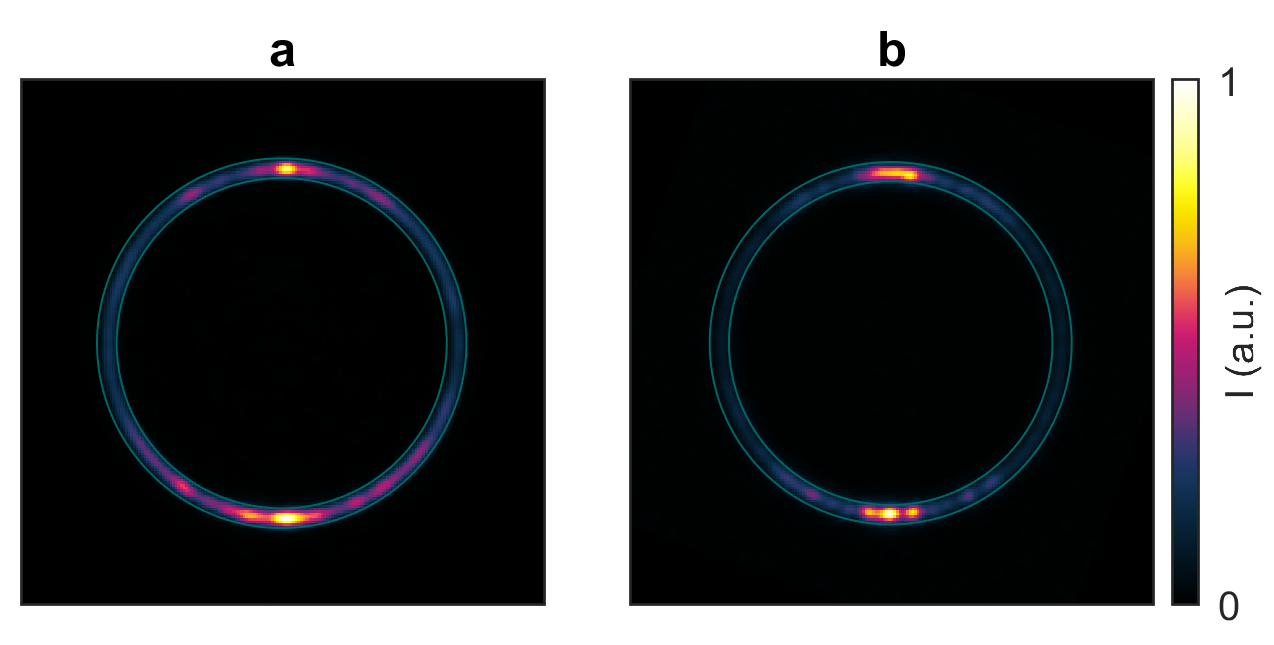}
    \caption{\textbf{Effect of the input polarization on the output intensity:} The angle-corrected averaged intensities for 60 different angular input positions of the light field for a fiber of half Talbot length with different input light field polarizations. The output intensity obtained using a linearly (a) and radially (b) polarized input light field.}
    \label{fig:ComparePol}
\end{figure}

The cylindrical self-imaging in a weakly-guiding ring-core fiber requires a radially polarized input light field. For this reason, we use an S-waveplate in the experiment, which ensures a radial polarization for all angular input positions.
To demonstrate the effect of the polarization on the self-imaging, we perform measurements using a fiber of half Talbot length and a single Gaussian spot on the input facet of the fiber resulting in two output spots. The measurement is done for linearly and radially polarized light fields, and 60 evenly distributed angular input positions.
In Fig.\,\ref{fig:ComparePol} (a), we show the angle-corrected average intensity over all input positions for a linearly polarized light field (here vertically polarized), and similarly for a radially polarized light field
in (b).
Although the expected pattern is visible in both cases, the self-imaging quality is clearly enhanced when using radially polarized light fields.

\bibliography{references}